\documentclass[journal,twoside,web]{ieeecolor}
\usepackage{tuffc}
\usepackage{cite}
\usepackage{amsmath,amssymb,amsfonts}
\usepackage{algorithmic}
\usepackage{graphicx}
\usepackage{textcomp}
\usepackage{wrapfig,colortbl}
\definecolor{abstractbg}{rgb}{1,0.969,0.914}
\usepackage{tcolorbox}

\DeclareMathOperator{\E}{\mathbb{E}}
\DeclareMathOperator{\R}{\mathbb{R}}

\newcommand{\ts}{{t_s}}
\newcommand{\bw}{{\mathbf w}}
\newcommand{\bd}{{\mathrm d}} 
\newcommand{\bx}{{\mathbf x}}
\newcommand{\bbx}{{\bar{\mathbf{x}}}}
\newcommand{\by}{{\mathbf y}}
\newcommand{\bh}{{\mathbf h}}

\newcommand{\bI}{{\mathbf I}}

\def\BibTeX{{\rm B\kern-.05em{\sc i\kern-.025em b}\kern-.08em
    T\kern-.1667em\lower.7ex\hbox{E}\kern-.125emX}}
\begin{document}
\title{Active inference and deep generative modeling for cognitive ultrasound}
\author{Ruud JG van Sloun \IEEEmembership{Member, IEEE}
\thanks{\textcopyright 2024 IEEE. DOI: 10.1109/TUFFC.2024.3466290. Submitted on July 5, 2024; accepted on September 16, 2024. This work was supported by the European Research Council (ERC) under the ERC starting grant nr. 101077368 (US-ACT), and the Dutch Research Council (NWO) under VIDI grant nr. 20381}
\thanks{Ruud JG van Sloun is with the Eindhoven University of Technology, Eindhoven, The Netherlands (e-mail: r.j.g.v.sloun@tue.nl)}}

\IEEEtitleabstractindextext{%
\fcolorbox{abstractbg}{abstractbg}{%
\begin{minipage}{\textwidth}\rightskip2em\leftskip\rightskip\bigskip
\begin{wrapfigure}[13]{r}{3in}%
\hspace{-3pc}\includegraphics[width=2.9in]{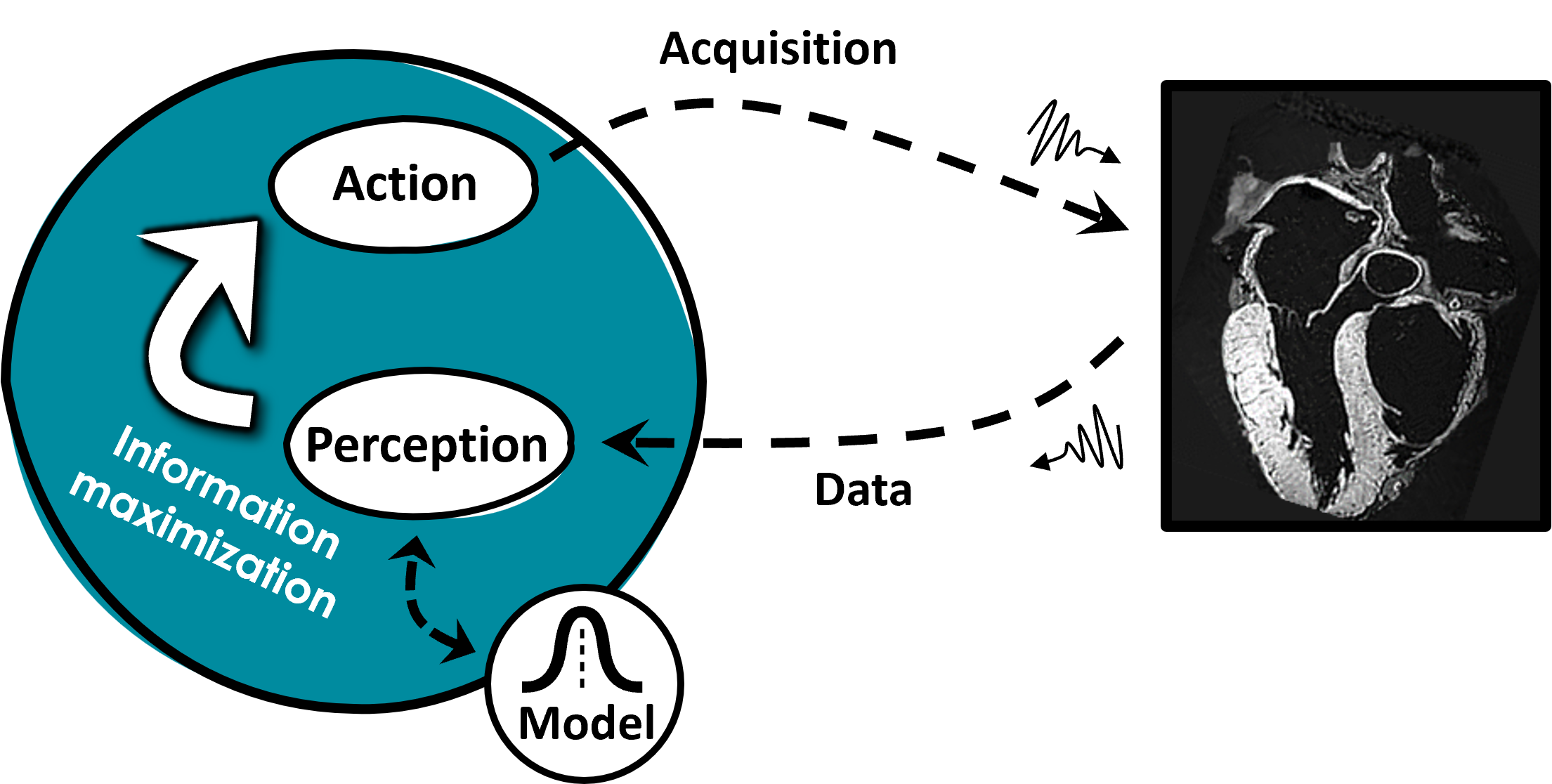}
\end{wrapfigure}%
\begin{abstract} 
Ultrasound has the unique potential to offer access to medical imaging to anyone, everywhere. Devices have become ultra-portable and cost-effective, akin to the stethoscope. Nevertheless, and despite many advances, ultrasound image quality and diagnostic efficacy are still highly operator- and patient-dependent. In difficult-to-image patients, image quality is often insufficient for reliable diagnosis. 
In this paper, we put forth the idea that ultrasound imaging systems can be recast as information-seeking agents that engage in reciprocal interactions with their anatomical environment. Such agents autonomously adapt their transmit-receive sequences to fully personalize imaging and actively maximize information gain in-situ. To that end, we will show that the sequence of pulse-echo \textit{experiments} that an ultrasound system performs can be interpreted as a perception-action loop: the action is the data acquisition, probing tissue with acoustic waves and recording reflections at the detection array, and perception is the inference of the anatomical and or functional state, potentially including associated diagnostic quantities. We then equip systems with a mechanism to actively reduce uncertainty and maximize diagnostic value across a sequence of experiments, treating action and perception jointly using Bayesian inference given generative models of the environment and action-conditional pulse-echo observations. Since the representation capacity of the generative models dictates both the quality of inferred anatomical states and the effectiveness of inferred sequences of future imaging actions, we will be greatly leveraging the enormous advances in deep generative modelling (\textit{generative AI}), that are currently disrupting many fields and society at large. Finally, we show some examples of cognitive, closed-loop, ultrasound systems that perform active beamsteering and adaptive scanline selection, based on deep generative models that track anatomical belief states. 
\end{abstract}

\bigskip
\end{minipage}}}

\maketitle

\section{Introduction}
\label{sec:introduction}

\IEEEPARstart{U}{ltrasound} (US) has the potential to revolutionize and democratize medical imaging due to its cost-effectiveness and portability. However, achieving consistent, precise and robust diagnostics remains a challenge. The diagnostic performance of US is dependent on skilled operators and exams still fail frequently on hard-to-image patients. The group of hard-to-image patients is moreover growing rapidly due to the rising incidence of obesity worldwide. Studies show that reduced image quality leads to worse observer variability, reproducibility, and accuracy of diagnostic parameters in ultrasound exams \cite{cole2015defining,nagata2018impact}. 

The biggest adversaries for ultrasound image quality and diagnostic accuracy stem from patient- and user-specific factors, i.e. patient geometry and user interaction. These factors thus vary across exams, and within exams. Given this, it is reasonable to hypothesize that optimal imaging requires closed-loop, goal-directed system behaviour, through sequential optimal experiment (=transmit-receive) design via its reciprocal interactions with the physical environment.   

Based on this hypothesis, this paper proposes a brain-inspired paradigm for such a \textit{cognitive} ultrasound transmit-receive control system. It recognizes that the cycle of ultrasound data acquisition and reconstruction can be interpreted as a perception-action loop: the action is the acquisition, probing the anatomy, and the perception is the reconstruction that infers what object most likely generated that acquired data. This data acquisition cycle has associated costs (e.g. time and energy), and hence in practice one always deals with partial observations of the time-varying object. 

Perception-action loops are commonly used in neuroscience and cybernetics to explain the behaviour of intelligent agents, which also typically deal with partial observations of the world around them. A now widely-accepted brain theory is that agents establish an internal generative model of the world in order to efficiently infer causes of their sensations, and plan useful future actions, all driven by an intrinsic motivation \cite{biehl2018expanding}: minimization of uncertainty. 

Through this lens, we will interpret the ultrasound imaging system as an agent, more specifically an \textit{active perceiver} \cite{bajcsy1988active, bajcsy2018revisiting}, which strives to perform actions and perceptions that minimize uncertainty about the anatomical world across a sequence of imaging experiments and resulting observations. To be effective at achieving this, the agent must understand the environment, and the consequences that actions have on observations of that environment. We postulate that this can be achieved by equipping the agent with generative models that govern these beliefs. Through probabilistic inference, an agent can then optimally plan its measurements, autonomously seek value, and continuously update its imaging strategy based on the incoming sensor data. In effect, full system behaviour is governed by the single holistic information-theoretic objective to minimize uncertainty. 

The above ambition becomes practical only when the agent's generative model is sufficiently expressive to reflect diverse, yet plausible, anatomical states and the intricacies of the relationship between an anatomical state, transmit-receive parameters (e.g. transmit waveforms or receive compression/sampling), and the observations. This is not at all trivial. Anatomical states (e.g. reflectivity) are typically represented on a spatial grid of tens- to hundreds of thousands of pixels, and the relationship between these pixels is intricate, with structure at various hierarchical scales. Additionally, there is structure in time, again at various scales. Moreover, since the agents' actions are determined by evaluating some expected observational value functional across hypotheses about the current (and future) state, it is critical that these hypotheses are not only plausible and consistent (i.e. they agree with the observations done thus far), but that they cover all modes of the true distribution to prevent collapse of the system into a degenerate, ``ignorant'', mode. 

Deep generative models, such as normalizing flows and diffusion models, have revolutionized generative modeling on all these aspects (scale, plausibility, and diversity) in past years. They are revolutionary in terms of sample fidelity and have already been used effectively to solve challenging image-based inverse problems. We are currently seeing the very first use cases for such models in ultrasound image reconstruction \cite{stevens2024dehazing, zhang2023ultrasound, lan2023fast, asgariandehkordi2023deep}, and will here also elucidate their potential in the context of active inference and closed-loop cognitive ultrasound. 

The remainder of this paper is organized as follows. In sections~\ref{sec:PALoops} and \ref{sec:DGM} we will introduce perception-action loops, and the rationale for using deep generative models, respectively. We will give some special attention to diffusion models. Then in section~\ref{sec:perceptual_inference}, we will give a formal description of what exactly we mean by ``perceptual inference'', the rationale for choosing a Bayesian approach, and various approaches to executing this tractably. We will then, in section~\ref{sec:active_inference}, dive into the inference of actions (the active part of {active inference}), the design of criteria that measure the value of actions, and mechanisms to evaluate \textit{expected future value} based on generative predictions of ``what may happen''. Section~\ref{sec:approximation} gives an overview of methods for approximating the generative density functions as well as the action-value functions in complex models when exact computation is intractable. Then, in section~\ref{sec:examples}, we will give some concrete examples that have tutorial value, illustrating the potential utility of the presented approaches in the context of ultrasound. Finally, in section~\ref{sec:discussion}, we will discuss future research directions, open questions, and outstanding challenges, and then conclude in section~\ref{sec:conclusion}. {Table~\ref{table} gives a glossary of some terms and symbols that are used throughout this paper.} 

\vspace{-0.3cm}
\begin{table}[h]
\label{table}
\centering
\caption{Glossary of terms and symbols used in this paper.}
\begin{tabular}{|p{2cm}|p{6cm}|}
\hline
\textbf{Term} & \textbf{Definition} \\ \hline
$x, y, a$ & Random variables. \\ \hline
$x', y', a'$ & Deterministic variables. \\ \hline
$\hat{y}, \hat{a}$ & Data points. \\ \hline
$x^i, y^i$ & Samples from a probability density function. \\ \hline
$p(x,y,a)$ & A (probabilistic) generative model across states, observations and actions. We will use both generative models based on first principles, and deep generative models, learned from data. \\ \hline
$p(x, y, a')$ & A generative model across states and observations evaluated for a particular deterministic action $a'$. Shorthand notation for $p(x, y, a=a')$. \\ \hline
$p(x|\hat{y}, \hat{a})$ & A Bayesian posterior distribution for the states given past and present observations and actions. Shorthand notation for $p(x|y=\hat{y}, a=\hat{a})$. \\ \hline
$q(x|\hat{y}, \hat{a})$ & An approximate posterior distribution. \\ \hline
$\E_{q(x)}f(x)$ & The expected value of function $f(x)$ when $x\sim q(x)$. \\ \hline
$I(x, y| a')$ & The mutual information between state $x$ and observation $y$, for a particular action $a'$. Reflects the information gain of an experiment with action $a'$. \\ \hline
$H(y|a')$ & The marginal entropy of observations $y$, for a particular action $a'$. Reflects the uncertainty about outcomes of observations $y$ for an action $a'$. \\ \hline
$H(y|x, a')$ & The conditional entropy of $y$ given $x$, for a particular action $a'$. Equal to the expectation $\E_{x'\sim q(x)}H(y|x=x',a=a')$. Reflects the remaining uncertainty about $y$ when given the state $x$.\\ \hline
$s_\theta(x)$ & A neural-network-based approximation of the true score function of a probability distribution $p(x)$: $\nabla_{x}\log p(x)$. Used in diffusion models to draw samples from the data distribution. \\ \hline
$|\Sigma|$ & Determinant of a covariance matrix $\Sigma$, also referred to as the generalized variance. \\ \hline
\end{tabular}
\end{table}
\vspace{-0.2 cm}
\section{Perception-action loops}
\label{sec:PALoops}
\noindent \textit{“Each movement we make by which we alter the appearance of objects should be thought of as an experiment designed to test whether we have understood correctly the invariant relations of the phenomena before us”
										— Helmholtz \cite{hermann1971selected}}
\\ \\
Intelligent agents, such as the brain, continuously engage in interactions with their environment. These interactions are reciprocal, i.e. agents take actions which affect their environment, and their environment in turn affects their observations. These observations solicit new actions, closing the so-called \textit{perception-action loop}. Useful actions are those that lead to desired outcomes/observations. To be effective at achieving this, an agent must understand the environment, and the consequences of actions on that environment. Rational agents thus pursue the understanding of their environment through actions that lead to information gain, while at the same time being goal-directed. Such behavior requires the ability to make predictions about the environment, and thus agents embody a generative model. 

This paper will be concerned only with perception-action loops in which actions influence the observation/measurement of an environmental state, but not the state itself. The agent is thus an active perceiver. As we will see, this has some implications for the way we factorize generative models and their approximate posteriors. Some of the assumptions and parameterizations used here will therefore deviate from what is typical in the control/cybernetics literature, which assumes actions impact states of the environment, rather than observations thereof. We will further restrict ourselves to agents that have no explicit preferences about outcomes, other than to seek information gain. They are ``scientists'' planning successive experiments to better understand what phenomenon causes their observations.
\vspace{-0.1cm}
\begin{tcolorbox}[title={The ultrasound action space}]
At this point, it becomes useful to briefly discuss what the action space of an ultrasound agent entails. Many of the controllable sensing parameters in ultrasound are similar to those commonly considered in cognitive radar systems. They include both transmit and receive parameters. On the transmit side, at the most abstract level, the agent is tasked with the design of an optimal, maximally informative, transmit code. At the most fine-grained level, this code is governed by a complete shaping of the transmit waveforms and timing for each of the transmitters, subject to the constraints of feasibility imposed by the analog transmit chain. A more coarse representation would be to only control the delays and gains applied to each transmitter, shaping the transmit beam. Practical ultrasound systems typically execute a sequence of such coded transmit events (e.g. a series of focused scan lines), and many practical codes will need to be considered in the context of a full sequence to evaluate the impact they have on the information gain achieved by that sequence. 
\end{tcolorbox}
\vspace{-0.1cm}

\begin{figure}
    \centering
    \includegraphics[width=1.08\linewidth, trim={1cm 11cm 18cm 0},clip]{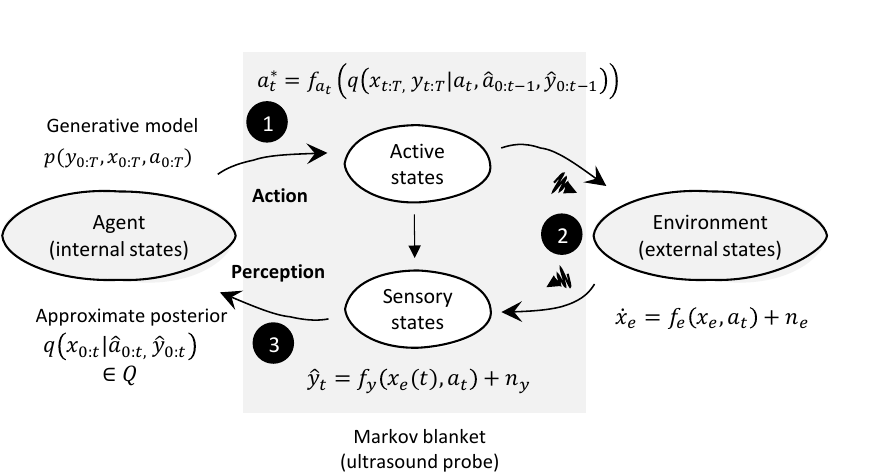}
    \caption{At time point $t$, an agent equipped with a generative model $p$ selects an action (1), which manifests in an excitation of the environment (2). The excitation ``changes'' the environment (e.g. it introduces compressional waves). This in turn results in a new sensory state $\hat{y}_t$. Confronted with the updated sensory data $\hat{y}_{0:t}$, the agent then revisits its beliefs about the environment (including future states it may take, and observations that may follow from that), and computes a new (approximate) posterior $q$ (3). The ultrasound probe contains the active and sensory states, and acts as a Markov blanket that separates the agent from its environment; they only interact via the active and sensory states. The distinction between $x_e$, the environmental states, and $x$, the internal states, is to make explicit that the agent's model is in general an approximate model of the true physical environment.}
    \label{fig:markov_blanket}
    \vspace{-0.5cm}
\end{figure}

We define the following time-discretized perception-action loop. At time point $t$, an agent equipped with a generative model selects an action, which manifests in an excitation of the environment. This in turn results in a new sensory state $\hat{y}_t$. Confronted with the updated sensory data $\hat{y}_{0:t}$, the agent then revisits its beliefs about the environment (including future states it may take, and observations that may follow from that), and computes a new posterior belief about the state. Figure~\ref{fig:markov_blanket} gives an overview of the perception-action loop. 

The agent's generative model over observations $y_{0:T}$,  environmental states $x_{0:T}$, and actions $a_{0:T}$ is given by:
\begin{equation}
    \label{eqn:genmodel}
    y_{0:T}, x_{0:T}, a_{0:T} \sim p(y_{0:T},x_{0:T},a_{0:T}),
\end{equation}
with
\begin{align}
    \label{eqn:genmodel2}
p(y_{0:T},x_{0:T},a_{0:T}) &= p(y_{0:T}|x_{0:T}, a_{0:T})p(x_{0:T},a_{0:T}) \nonumber \\
 &= p(y_{0:T}|x_{0:T}, a_{0:T})p(x_{0:T})p(a_{0:T}|x_{0:T}),
\end{align}
by which we make explicit that actions impact observations, not states (active perception), and that the state impacts the distribution across actions (adaptivity). In section~\ref{sec:active_inference} we will show how the latter dependency manifests precisely in our framework, but in a nutshell it comes down to the maximization of an information-theoretic action-value functional of the \textit{conditional} generative model $p(y_{0:T},x_{0:T}|a_{0:T})$. Note that the active-perception model foregoes that in reality, actions do impact the physical environment (they introduce compressional waves), and considers all these effects part of the observation model $p(y_{0:T}|x_{0:T}, a_{0:T})$. This distinction between the agent's internal model, and the true environment is also made explicit in Figure~\ref{fig:markov_blanket}.

Throughout this tutorial, we also refer to the generative models given above as \textit{priors}. Whenever these prior beliefs are revised through exposure to information in the form of data/measurements, we refer to the revised beliefs as \textit{posteriors}. We refer to the updating of beliefs in the face of data as perceptual inference, which will be treated in section~\ref{sec:perceptual_inference}. Developing a sufficiently accurate generative model that enables reasoning about futures with the detail and diversity needed for fulfilling the agents' objectives is key. Our agent is tasked with inferring high-dimensional, high-resolution images, for which the requirements are particularly challenging. We will now briefly discuss how recent developments in deep generative modelling alleviate some of these challenges, offering new opportunities for active inference in the context of high-dimensional data.   

\begin{figure*}
    \centering
    \includegraphics[width=0.7\linewidth, trim={0 12cm 18cm 1cm},clip]{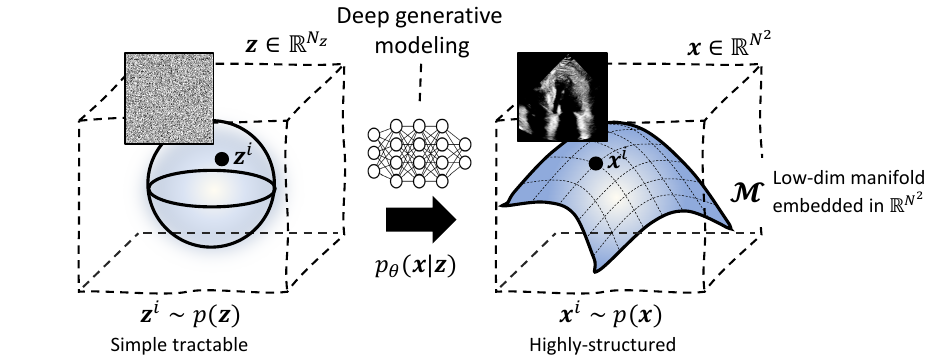}
    \caption{Real-world high-dimensional data such as ultrasound images lie on a low-dimensional manifold embedded in that high-dimensional space. This manifold is typically very intricate and non-smooth in the high-dimensional data space, and images that lie on it are highly structured. Deep generative learning allows modeling of such highly-structured distributions and sampling of novel datapoints that lie on these low-dimensional manifolds. This is enabled by transforming samples from a tractable distribution (such as an isotropic Gaussian) $z^i\sim\R^{N_z}$ into samples from the true data distribution $x^i\sim\R^{N^2}$. There are many ways of achieving this, e.g. via a conditional distribution $p_\theta(x|z)$ trained using variational inference, game theory (adversarial models), or just maximum likelihood for specific invertible models (normalizing flows). Alternatively, iterative sampling methods learn to estimate the gradients of the true data distribution at a plurality of noise scale manifolds that successively corrupt the data distribution into a tractable isotropic normal. These gradients then allow for reversing this process using reverse diffusion or Langevin dynamics.}
    \label{fig:manifolds}
\end{figure*}

\section{Deep generative models}
\label{sec:DGM}
\noindent The potential gains of using accurate generative priors in ultrasound perception-action loops can perhaps most easily be understood when realizing that most combinations of pixels that form images are not interesting at all. Although a 5-second high-quality ultrafast video can in principle represent more than a trillion unique instances, the vast majority of those hypothetical instances is unstructured and unnatural. If one were to completely at random draw an instance vector $x \in \R^{N_x \times N_y \times N_t}$, the resulting sample would with overwhelming probability be classified as ``noise''. In fact, one has to draw an extremely large amount of samples in this fashion (with this amount scaling very quickly in the number of dimensions), until any ultrasound image is sampled. Only a tiny fraction of that space is occupied by plausible ultrasound images. This is the manifold hypothesis: real-world high-dimensional data (such as images) lie on low-dimensional manifolds embedded within the high-dimensional space \cite{fefferman2016testing}. To use generative models in perception-action loops is to use the structure of the natural world. It allows systems to track states on a low-dimensional manifold, rather than in the very high-dimensional data space. Deep generative models can be used to accurately describe the complex hierarchical structure in spatio-temporal imaging data (see Fig.~\ref{fig:manifolds}). 

In principle one could choose to learn the entire generative model in Eqn.~\eqref{eqn:genmodel} from data. Given a training dataset of $L$ samples $\{(\hat{y}_{0:T}^1,\hat{x}_{0:T}^1,\hat{a}_{0:T}^1), ... , (\hat{y}_{0:T}^L,\hat{x}_{0:T}^L,\hat{a}_{0:T}^L) \}$, one would then fit a deep generative model to the joint distribution. Often we will use the factorization in Eqn.~\eqref{eqn:genmodel2}, and when possible make use of modelling and physics to describe the conditional observation density $p(y_{0:T}|x_{0:T},a_{0:T})$, which we will often be able to further decompose into memoryless factors $p(y_t|x_t,a_t)$. Many observation models are reasonably assumed to be Gaussian, with its mean being some (possibly nonlinear) function of the state $x_t$ and action $a_t$. This leaves us with the modelling of the prior $p(x_{0:T})$

How much structure (in space and time) should be imposed on $x_{0:T}$ amounts to the classic bias-variance trade-off. One way to reduce bias is to factorize the prior, e.g. by only imposing structure locally in space or time (patches). That is, we assume that there is a factorization
\begin{equation}
\label{eqn:factorize}
p(x) = \prod_i p(x_i),
\end{equation}
with factors $p(x_i)$ governing the density functions for non-overlapping individual patches $x_i\in \R^{M_x\times M_y\times M_t}$. 

Given this, we can now fit a deep generative model for $p(x)$ to its data samples. Deep generative modelling comes in many flavours, and which technique to use is a design choice. We refer the reader to the book by Tomczak \cite{tomczak2021DGM} for an excellent overview. In this paper, we are mostly concerned with the generation of samples, i.e. $x^i\sim p(x)$, where $p(x)$ is some complex data distribution (e.g. that of images). This is enabled by drawing samples from some tractable, simple distribution $z^i\sim p(z)$, and transforming those into samples from $p(x)$. Using variational inference \cite{kingma2013auto} one can optimize an evidence lower bound to $p(x)$ by jointly learning the generative conditional distribution $p_\theta(x|z)$ and a variational posterior $q_\phi(z|x)$ that is amortized across the training dataset, given a tractable prior e.g. $p(z)=\mathcal{N}(0,I)$. Another alternative is to use game theory and train models that play a game against a neural adversary trying to detect whether generated samples come from the true data distribution or are generated by the model \cite{creswell2018generative}.  We can also train models directly using maximum likelihood when using specific invertible models, i.e. normalizing flows \cite{kobyzev2020normalizing}. Iterative sampling methods instead learn to estimate the gradients of the true data distribution at a plurality of noise scale manifolds that successively corrupt the data distribution into a tractable isotropic normal. These gradients then allow for reversing this process using reverse diffusion methods or Langevin dynamics. We will now briefly review diffusion models as they will later be used in our examples. \\

\noindent\textbf{Diffusion models:}
Diffusion models are state-of-the-art in image and video generation. A notable challenge that diffusion models overcome is the need for explicit normalization of learned density functions (i.e. assuring that $\int p(x) \bd x = 1$). Such normalization is typically achieved by imposing specific constraints on the architecture (e.g., using flows \cite{kobyzev2020normalizing}) or using variational approximate inference methods. Instead, diffusion models indirectly parameterize the (log) data distribution by learning to approximate its gradient, the score function.

The score function appears when reversing a forward stochastic differential equation (SDE) that progressively adds Gaussian white noise to samples drawn from the data distribution $p(\bx)$, such that eventually these samples become samples from a standard Normal distribution \cite{song2020score}. The {  forward} SDE is as follows:
\begin{equation}
    \bd x = - \frac{\beta(\ts)}{2}x \bd\ts + \sqrt{\beta(\ts)}\bd w,
\end{equation}
where $x(0)$ is an initial noise-free sample, $\ts \in [0, \mathcal{T}]$, $\beta(\ts)$ is the noise schedule, $\bw$ is a standard Wiener process, and $x(\mathcal{T}) \sim \mathcal{N}(0, I)$. As said, this SDE can be reversed using the score function \cite{anderson1982reverse}:
\begin{equation}
\label{eq:reverse-diffusion}
    \bd x = \left[ -\frac{\beta(\ts)}{2}x - \beta(\ts)\nabla_{x}\log p_\ts (x)\right]\bd\ts + \sqrt{\beta(\ts)}\bd\bar{w},
\end{equation}
where $\bar{w}$ is a standard Wiener process running backwards. Following the notation by \cite{chung2022diffusion}, the discrete setting of the SDE is represented using $x_\ts = x(\ts T/N), \beta_\ts = \beta(\ts T/N), \alpha_\ts = 1-\beta_\ts, \bar{\alpha}_\ts = \prod_{s=1}^\ts\alpha_s$.

The diffusion model performs reversal of the forward SDE by learning the score function using a neural network parameterised by $\theta$, conditioned on the timestep $\ts$, $s_\theta(x_\ts, \ts) \approx \nabla_{x}\log p_\ts (x)|_{x=x_\ts}$. The parameters are learned using denoising score matching \cite{songScoreBasedGenerativeModeling2021}.
To sample from $p(x)$, one can then sample from a standard normal, and revert the SDE using the learned score function. This is an iterative process, with a neural network function evaluation at each iteration, and hence its vanilla implementation is time-consuming. Accelerating diffusion models is hence a very active research theme today, with many solutions for fast sampling being developed \cite{meng2023distillation}.

\section{Perceptual inference}
\label{sec:perceptual_inference}
\noindent We will now make the perception step in our perception-action loop explicit. \textit{Perception} is the act of updating beliefs about the states governed by the generative model in Eqn.~\eqref{eqn:genmodel} given new information. Indeed, information is that which induces a change of beliefs. Note that this is distinct from \textit{learning}, which is to infer the generative model (functional form and/or parameters) itself and occurs across longer timescales. In the following, we will assume that the agent has established or has access to a sufficiently accurate generative model. This may be obtained through past interactions and observations (training data) and/or injected knowledge of physics, that is available at $t=0$. 

At time step $t$ the agent first takes an \textit{action} $\hat{a}_t$, which leads to observations $\hat{y}_t$. Given the tuple of all observations and actions performed thus far, which we denote as { $\hat{\bar{y}}_{0:t} = (\hat{y}_{0:t}, \hat{a}_{0:t})$}, a rational agent will seek the minimal update of past/prior beliefs about $x_{0:t}$ that satisfies the constraints imposed by the new data - beliefs must only be revised to the extent required by the new information \cite{jaynes1982rationale, caticha2011entropic}. Such belief updates in the face of data are given by Bayes rule:
\begin{equation}
p(x_{0:t}|\hat{\bar{y}}_{0:t}) = \frac{p(\hat{\bar{y}}_{0:t},x_{0:t})}{p(\hat{\bar{y}}_{0:t})} = \frac{p(\hat{\bar{y}}_{0:t}|x_{0:t})p(x_{0:t})}{p(\hat{\bar{y}}_{0:t})}.
\end{equation}
The numerator can be evaluated using the known generative model Eqn.~\eqref{eqn:genmodel}. Unfortunately, computing the marginal $p(\hat{\bar{y}}_{0:t})=\int p(\hat{\bar{y}}_{0:t}|x_{0:t})p(x_{0:t})dx_{0:t}$ quickly becomes infeasible for high-dimensional states. In practice we will therefore typically approximate the true posterior distribution, using an \textit{approximate posterior} $q(x_{0:t}|\hat{\bar{y}}_{0:t})$. We will elaborate on this in section~\ref{sec:approximation}. 

Perceptual inference also revisits hypotheses about what has yet to come. That is, our posterior beliefs about the past and present yield updated priors about the future. { Using the generative model in Eqn.~\eqref{eqn:genmodel2}, and assuming that (i) at timestep $t$, the actions $\hat{a}_{0:t}$ and observations $\hat{y}_{0:t}$ are known,  i.e. $q(y_{0:t},a_{0:t})=\delta(y_{0:t}-\hat{y}_{0:t})\delta(a_{0:t}-\hat{a}_{0:t})$, (ii) actions only influence observations and not states (active perception), and (iii) $y_{t}$ is only a function of $x_{t}$ and not of past states (the observation model is memoryless), the approximate posterior projected also into \textit{future} states and observations, for a deterministic sequence of hypothetical \textit{future} actions $a'_{t+1:T}$, is given by:
\begin{align}
\label{eqn:approxpost}
    q(x_{0:T}, \hspace{3pt} & y_{t+1:T}| a'_{t+1:T}, \hat{\bar{y}}_{0:t}) \nonumber \\  \overset{(ii)}{=}& p(y_{t+1:T}|x_{0:T},a'_{t+1:T})q(x_{0:T}|\hat{\bar{y}}_{0:t}) \nonumber \\
    \overset{(iii)}{=}& p(y_{t+1:T}|x_{t+1:T},a'_{t+1:T})q(x_{0:T}|\hat{\bar{y}}_{0:t}) \nonumber \\
    =& \underbrace{p(y_{t+1:T}|x_{t+1:T},a'_{t+1:T})}_{\text{observation model}}\underbrace{p(x_{t+1:T}|x_{0:t})}_{\text{state evolution}}\underbrace{q(x_{0:t}|\hat{\bar{y}}_{0:t})}_{\text{current posterior}}.
\end{align}
Note the explicit distinction between known forward generative processes, indicated by $p(\cdot)$, and approximations of the exact Bayesian posteriors under the generative process, indicated by $q(\cdot)$.}

Equation~\eqref{eqn:approxpost} enables the generation of hypothetical state-observation pairs in the future, which are revisited \textit{in-situ} as new data $\hat{\bar{y}}$ comes in:
\begin{align}
q(& x_{t+1:T}, y_{t+1:T} | a'_{t+1:T}, \hat{\bar{y}}_{0:t})\nonumber \\ & \hspace{5pt} = \E_{q(x_{0:t}|\hat{\bar{y}}_{0:t})} p(y_{t+1:T}|x_{t+1:T},a'_{t+1:T})p(x_{t+1:T}|x_{0:t}).
\end{align}

We will now proceed with the selection of information-optimal actions based on these hypothetical futures. 

\section{Active inference and information gain}
\label{sec:active_inference}
\noindent \textit{“What is the first and most fundamental thing a newborn infant has to do? If one subscribes to the free energy principle, the only thing it has to do is to resolve uncertainty about causes of […] sensations.” 
									– Friston \cite{friston2017self}}
\\ \\
The active inference paradigm postulates that an agent's desires are encoded by their generative models, which attribute more mass to states and observations that are aligned with preferred outcomes \cite{friston2016active }. Humans desire to stay alive and hence place a strong prior on equilibrium conditions manifesting in e.g. homeostasis, and subsequent observations such as ``not being hungry or too cold''. In active inference, these priors subsequently drive inference of actions through the pursuit of observations and rational beliefs about hidden states that are not surprising. In doing so, agents actively resist the increase of posterior entropy. 

As mentioned before, we will here restrict our treatment to agents that take actions that are purely ``scientific'' in nature, i.e. they maximize information gain, placing no explicit utilitarian preference on particular observations. We will discuss agents that do not strive for survival but ``just'' seek to understand the world from the viewpoint of an external observer \cite{biehl2018expanding}.   

We start by defining the action-value functional, $V(\cdot)$, which evaluates the expected future value for each sequence of actions $a'_{t+1:T}$ given a probability density function governing our present beliefs about future states and observations $V(a'_{t+1:T} ; q(x_{t+1:T},y_{t+1:T} | a'_{t+1:T}, \hat{\bar{y}}_{0:t}))$. For brevity, we will let the conditioning on the probability density function be implicit in what follows. We will also use the shorthand notation $\tau = t+1:T$. We seek to minimize uncertainty about the state $x_{\tau}$, and hence our action-value-functional of choice will be the negative expected posterior entropy:
\begin{align}
\label{eqn:H(x|y)}
V(a'_{\tau}) \triangleq&- H(x_{\tau }|y_{\tau },a'_{\tau},\hat{\bar{y}}_{0:t}) \\ =& \E_{q(x_{\tau},y_{\tau}|a'_{\tau},\hat{\bar{y}}_{0:t})}\log \frac{q(x_{\tau},y_{\tau}|a'_{\tau},\hat{\bar{y}}_{0:t})}{q(y_{\tau}|a'_{\tau},\hat{\bar{y}}_{0:t})}, \nonumber
\end{align}
and the optimal action is:
\begin{align}
a_{\tau}^* = \underset{a'_{\tau}}{\arg\max} \hspace{5pt} V(a'_{\tau}).
\end{align}

Colloquially, the system will pursue actions that are expected to ``sharpen’’ the posterior the most. 
As was already shown by Lindley \cite{lindley1956measure}, minimizing expected posterior entropy is equivalent to maximizing the conditional mutual information between state and future observations, such that we have:
\begin{align}
a_{\tau}^* =& \underset{a'_{\tau}}{\arg\max} \hspace{5pt} I(x_{\tau},y_{\tau}|a'_{\tau},\hat{\bar{y}}_{0:t}) \\
=& \underset{a'_{\tau}}{\arg\max} \underbrace{\E_{q(x_{\tau},y_{\tau}|a'_{\tau},\hat{\bar{y}}_{0:t})}\log p(y_{\tau}|x_{\tau},a'_{\tau})}_{-H(y_{\tau}|x_{\tau},a'_{\tau},\hat{\bar{y}}_{0:t})} \nonumber \\ & \underbrace{- \E_{q(y_{\tau}|a'_{\tau},\hat{\bar{y}}_{0:t})}\log q(y_{\tau}|a'_{\tau},\hat{\bar{y}}_{0:t})}_{H(y_{\tau}|a'_{\tau},\hat{\bar{y}}_{0:t})}.
\label{eqn:mutual}
\end{align}
{ Note that we again explicitly distinguish factors that are inferred based on past data (the approximate posteriors $q(\cdot)$) and forward generative models $p(\cdot)$.} The conditional mutual information can also be interpreted as the expected prior-posterior gain, i.e. the expected divergence between the prior and posterior, or colloquially, ``how much it has been updated''. The larger the update, the more information has been gained from the measurement. 

From Eqn.~\eqref{eqn:mutual} we can also appreciate that maximizing mutual information, or information gain, can be achieved by selecting the actions that maximize the marginal differential entropy of future observations $H(y_{\tau}|a'_{\tau},\hat{\bar{y}}_{0:t})$, if the conditional entropy of the observations given the state is not dependent on the actions, i.e. $H(y_{\tau}|x_{\tau}, a'_{\tau})=H(y_{\tau}|x_{\tau}),  \forall a_\tau'$. This is often a reasonable assumption for active perceivers and generally holds for observation models of the form $y_t=f(x_t,a_t)+n_t$ in which the remaining entropy in $y_t$ given full knowledge of $x_t$ is solely determined by the noise $n_t$, and thus independent of $a_t$.   

\section{Implementing perception-action loops}
\label{sec:approximation}
\noindent Estimating the entropies in Eqn.~\eqref{eqn:mutual} is not trivial in practice, especially for the flexible class of density functions needed to accurately describe high-dimensional images and their (possibly nonlinear) observations. It is worth reiterating that simple linear models based on members of the exponential family that do have closed-form expressions are insufficient for what we try to achieve. Instead, we hypothesise that performing approximate inference with accurate (deep) generative models is more fruitful than pursuing exact inference with overly simple models. This confronts us with the daunting task of performing reliable \textit{approximate} inference using highly nonlinear (deep) generative models. For the sake of brevity, in what follows we make the dependency of future observations on the actions $a'_\tau$ implicit and use shorthand notation $\hat{\bar{y}}$ to indicate the tuple of all past observations and actions $\hat{\bar{y}}_{0:t-1}$.

\subsection{Approximate posterior inference}
A common approach to approximating the posterior density function is to use a set of samples/particles $\{x_{0:T}^1,...,x_{0:T}^{N_p}\}\sim p(x_{0:T}|\hat{\bar{y}})$ and weights $\{w_1,...,w_{N_p}\}$ that are proportional to the probability of the sample belonging to the target distribution:
\begin{equation}
\label{eqn:approx_post_sampling}
q(x_{0:T}|\hat{\bar{y}}) = \sum_{i=1}^{N_p} w_i \delta(x_{0:T}-x_{0:T}^i),
\end{equation} 
with $\sum_i w_i = 1$. 
Sampling methods make no explicit assumptions about the functional form of the density functions and hence have low bias. This does come at the expense of relatively high variance, and resolving it requires paying a computational price, e.g. in terms of the number of samples required to get good (mode) coverage of the distribution. 

Many methods allow for effective and/or efficient sampling from a posterior. For instance, deep generative diffusion models enable posterior sampling by integrating a data-consistency/likelihood step into the reverse diffusion process \cite{chung2022diffusion}. Another alternative is to use classic sequential Monte-Carlo methods (e.g. particle filters) \cite{cappe2007overview}. These methods exploit the sequential structure of states to perform highly efficient tracking of the distribution under arbitrary (possibly non-differentiable) nonlinear forward models. Such forward models can be physics-based, or learned from data, e.g. a pre-trained deep generative latent variable model that fits $p(x_t)=\int_z p(x_t|z)p(z)dz$ using variational inference \cite{kingma2013auto}. An extensive review of methods for approximating posterior distributions given nonlinear high-dimensional models is beyond the scope of this paper. We refer the interested reader to \cite{chopin2020introduction, zhang2018advances, chung2022diffusion, stevens2023removing}. As before, we will however give some special attention to diffusion models. \\

\noindent\textbf{Diffusion posterior sampling:}
\label{sec:DPS}
The reverse diffusion process can be conditioned on a measurement obtained using a model $y|x \sim p(y|x)$ to produce samples from the posterior $p(x|y)$. This is done by changing the unconditional score function in \eqref{eq:reverse-diffusion} into a posterior score function $\nabla_{x_\ts}\log p_\ts (x_\ts|y)$. However, the posterior score is not trivial to evaluate, as refactoring it into a (data-conditional) likelihood score, $\nabla_{x_\ts}\log p_\ts(y | x_\ts)$, and the original (unconditional) prior score, requires computing the intractable likelihood of noise-perturbed states. This has led to various approximate guidance schemes \cite{chung2022diffusion, song2022pseudoinverse, rout2024solving}, most of which exploit Tweedie’s formula, which can be thought of as a one-step denoising process from $\ts \rightarrow 0$ using our trained diffusion model to estimate the true fully-denoised sample $x_0$ as follows:
\begin{align}
    \label{eq:tweedie}
    \hat{x}_0 =& \mathbb{E}[x_0|x_\ts] \nonumber \\ \approx& \frac{1}{\sqrt{\bar{\alpha}(\ts)}}(x_\ts + (1-\bar{\alpha}(\ts))s_\theta(x_\ts, \ts)).
\end{align}
Diffusion Posterior Sampling (DPS) uses \eqref{eq:tweedie} to approximate $\nabla_{x_\ts}\log p(y|x_\ts)\approx \nabla_{x_\ts}\log p(y|\hat{x}_0)$, which for (non)linear Gaussian likelihood models leads to a guidance gradient $\nabla_{x_\ts}||y - f(\hat{x}_0)||^2_2$ that can be straightforwardly evaluated. This measurement-guidance gradient step is then alternated with the standard reverse diffusion steps using the unconditional score.

\subsection{Approximate entropy models}
Given a set of posterior samples, the next step is to estimate the entropy terms in our action-value function. { Recall that our action-value function is the action-conditional mutual information between state and future observations, and decomposes into two entropies: The marginal entropy of the observations, $H(y_\tau|\hat{\bar{y}})$, and the state-conditional entropy of the observations, $H(y_\tau|x_\tau, \hat{\bar{y}})$ (see Eqn.~\eqref{eqn:mutual}). In the most general case, both depend on the action, although there are many practical examples for which the latter dependency is negligible (recall the discussion at the end of Section~\ref{sec:active_inference}).

We nevertheless start with the expected conditional entropy (the remaining uncertainty about $y_\tau$ if $x_\tau$ were known) as choosing a reasonable entropy model is more straightforward. Assuming a forward measurement model with additive Gaussian noise $n_\tau$, \textit{i.e.} $y_\tau=f(x_\tau,a_\tau)+n_\tau$, the conditional entropy model is also a Gaussian. Given posterior samples $\{x_\tau^1,...,x_\tau^{N_p}\}\sim p(x_\tau|\hat{\bar{y}})$ with uniform weights, we then have that:}
\begin{align}
H(y_\tau|x_\tau, \hat{\bar{y}}) =& \E_{q(x_\tau,y_\tau|\hat{\bar{y}})}-\log p(y_\tau|x_\tau) \\ 
=& -\E_{q(x_\tau|\hat{\bar{y}})}\E_{p(y_\tau|x_\tau)}\log p(y_\tau|x_\tau) \\ 
= & - \frac{1}{N_p}\sum_{i=1}^{N_p}\E_{p(y_\tau|x_\tau^i)}\log p(y_\tau|x_\tau^i) \\
\approx& \hspace{5pt}\text{constant} + \frac{1}{N_p}\sum_{i=1}^{N_p}\frac{1}{2}\log |\tilde{\Sigma}_{y_\tau}(x_\tau^i)| \\
\overset{*}{=}& \hspace{5pt} \text{constant} + \frac{1}{2}\log |\tilde{\Sigma}_{y_\tau}|,
\label{eqn:approx_expectation2}
\end{align}
with $\tilde{\Sigma}_{y_\tau}(x_\tau^i)$ being the sample covariance, estimated using $N_p$ samples from $\{y_\tau^1,...,y_\tau^{N_p}\}\sim p(y_\tau|x_\tau^i)$, and the equality in Eqn.~\eqref{eqn:approx_expectation2} being true if $\tilde{\Sigma}_{y_\tau}(x_\tau^i)=\tilde{\Sigma}_{y_\tau}$ for all $x^i_\tau$.

We now proceed with the marginal entropy $H(y_\tau|\hat{\bar{y}})$. { Colloquially, this reflects the diversity of futures that may be observed for each of the actions. Ultimately, only one observation will manifest, but it is the diversity in hypothetical future observations that measures the uncertainty that one has about the result of an imaging experiment.} Approximating it requires choosing a reasonable marginal model, which is not obvious. In general, the marginal distribution over future observations $p(y_\tau|\hat{\bar{y}})$ may be multimodal, as the posterior distribution $p(x_\tau|\hat{\bar{y}})$ is likely multimodal too for high-dimensional states and partial observations. Nevertheless, in the examples of section~\ref{sec:examples} we will see that approximating the marginal as a multivariate Gaussian is often a reasonable choice from a pragmatic perspective. 

\begin{figure*}
    \centering
    \includegraphics[width=1\linewidth, trim={1cm 8.5cm 5cm 0},clip]{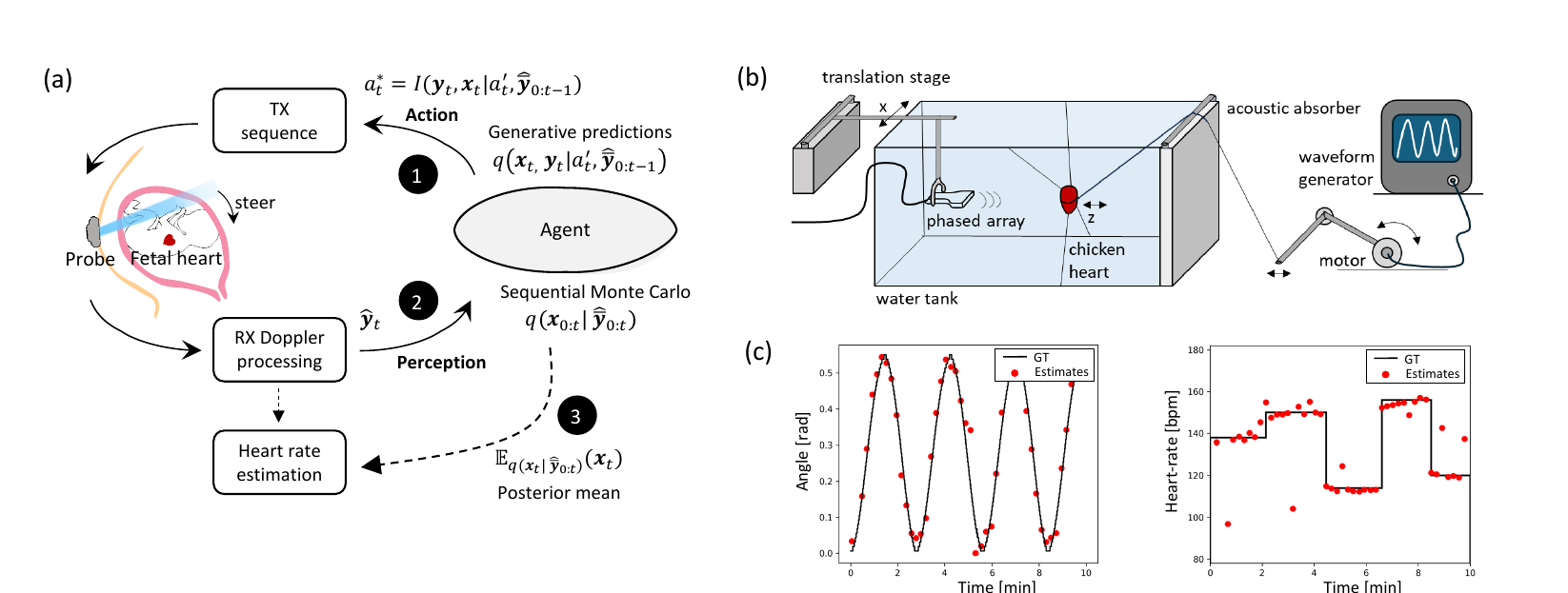}
    \caption{Example: Active beamsteering using sequential Monte-Carlo {\cite{federici2024active}}. (a) Doppler target tracking using cognitive ultrasound. (1) The agent selects the action (beamsteering angle $\theta_t^{\textrm{tx}}$) that has the highest expected information gain given generative predictions $q(\mathbf{x}_t,\mathbf{y}_t|a'_t,\hat{\mathbf{y}}_{0:t-1})$. (2) This action prompts a new Doppler observation $\hat{\mathbf{y}}_t$, which in turn triggers an update of the posterior $q(\mathbf{x}_{0:t}|\hat{\mathbf{y}}_{0:t})$, implemented using a sequential Monte Carlo method. (3) Finally, the posterior mean fetal heart location at that timestep is communicated to a heart rate estimation module alongside the received Doppler data. (b) Real-time lab setup mimicking the scenario described in (a), using a phased array transducer mounted on a translation stage that transmits a focused beam controlled by the agent to track a ``beating'' chicken heart. (c) Positional tracking and heart rate estimation (red) from adaptively steered focused beams, against ground truth (blue).}
    \label{fig:fetal}
\end{figure*}

{ To illustrate this, consider the following thought experiment. An agent evaluates the expected value of two candidate actions by sampling expected observations in $\R^1$ for both of them. The first action results in a unimodal empirical distribution over hypothetical observations of which all samples have some small nonzero $\ell_2$ distance to each other. The second action results in a bimodal distribution over hypothetical observations, with the $\ell_2$ distance among samples within each mode being close to zero, but the $\ell_2$ distance between the modes being very large. Fitting a (unimodal) Gaussian entropy model to both distributions would lead to a higher marginal entropy for the latter: the large $\ell_2$ distance between the modes would dominate the variance. As such the agent would select the latter action. In many risk-sensitive applications, such as medical imaging, selecting the latter action would arguably indeed be preferred, as it would discriminate between two equally plausible but (semantically) very different modes. The entropy of a Gaussian marginal is a reasonable surrogate for this ``distance'', although the model fit itself can be poor (as in the bimodal example given above).} The analogy is less intuitive for observations in $\R^N$, but it is clear that in that case one should also model the correlation across the elements in the observation vector.

Assuming that the marginal observations indeed follow a multivariate Gaussian pdf, i.e. $q(y_\tau|\hat{\bar{y}})=\mathcal{N}(\mu_{y|\hat{\bar{y}}},\Sigma_{y|\hat{\bar{y}}})$, and that we are given posterior samples $\{y_\tau^1,...,y_\tau^{N_p}\}\sim q(y_\tau|\hat{\bar{y}})$, we have:
\begin{align}
H(y_\tau|\hat{\bar{y}}) 
\approx & \hspace{5pt} \text{constant} + \frac{1}{2}\log |\tilde{\Sigma}_{y|\hat{\bar{y}}}|,
\end{align}
where $\tilde{\Sigma}_{y|\hat{\bar{y}}}$ is the sample covariance, estimated using $N_p$ samples from $p(y_\tau|\hat{\bar{y}})$. This result is intuitive: the determinant $|\tilde{\Sigma}_{y|\hat{\bar{y}}}|$ is also called the generalized variance, and is equal to the product of the eigenvalues of $\tilde{\Sigma}_{y|\hat{\bar{y}}}$.

Alternatively, we can start from the approximate posterior family in Eqn.~\eqref{eqn:approx_post_sampling} to derive a model for the marginal. Again, assuming Gaussian measurements $y_\tau = f(x_\tau, a_\tau) + n_\tau$, we obtain that $q(y_\tau|\hat{\bar{y}})$ is a mixture of Gaussians, i.e.:
\begin{align}
q(y_\tau|\hat{\bar{y}}) = \sum_{i=1}^{N_p} w_i \mathcal{N}(\mu_i,\Sigma_i),
\end{align}
where $\mu_i = f(x_\tau^i, a_\tau)$. A variational approximation to the marginal differential entropy is then given by \cite{hershey2007approximating}:
\begin{align}
\label{eqn:approxmixture}
H(y_\tau|\hat{\bar{y}}) \approx -\sum_{i=1}^{N_p} w_i \log \sum_{j=1}^{N_p} w_j e^{-KL[\mathcal{N}(\mu_i,\Sigma_i)||\mathcal{N}(\mu_j,\Sigma_j)]}\nonumber \\
+ \sum_{i=1}^{N_p} w_i H[\mathcal{N}(\mu_i,\Sigma_i)],
\end{align}
which can be evaluated in closed form. If one assumes that the measurement noise has constant diagonal isotropic covariance $\Sigma_i = \Sigma_j = \sigma_y^2 I$, Eqn.~\eqref{eqn:approxmixture} further simplifies to:
\begin{align}
\label{eqn:approxmixture2}
H(y_\tau|\hat{\bar{y}}) \approx \text{constant} - \sum_{i=1}^{N_p} w_i \log \sum_{j=1}^{N_p} w_j e^{-\frac{||\mu_i-\mu_j||_2^2}{2\sigma_y^2}},
\end{align}
which illustrates the intuitive connection between distance among the samples in observation space and the differential entropy.

We now have all the ingredients to close the ultrasound perception-action loop: (1) perceptual inference of anatomical state posteriors using deep generative models, followed by (2) action selection through maximization of an approximate action-conditional mutual information between future states and future observations.

\section{Examples}
\label{sec:examples}
\noindent { We will now illustrate the concepts described above in the context of concrete ultrasound imaging applications. From sections VII-A, to VII-C, we will cover (A) active inference with low-dimensional generative models based on first principles, (B) perceptual inference with complex data-driven deep generative models in high dimensions, and finally, (C) active inference using deep generative models.}

\begin{figure*}
    \centering
\includegraphics[width=1\linewidth, trim={0 9cm 11cm 0},clip]{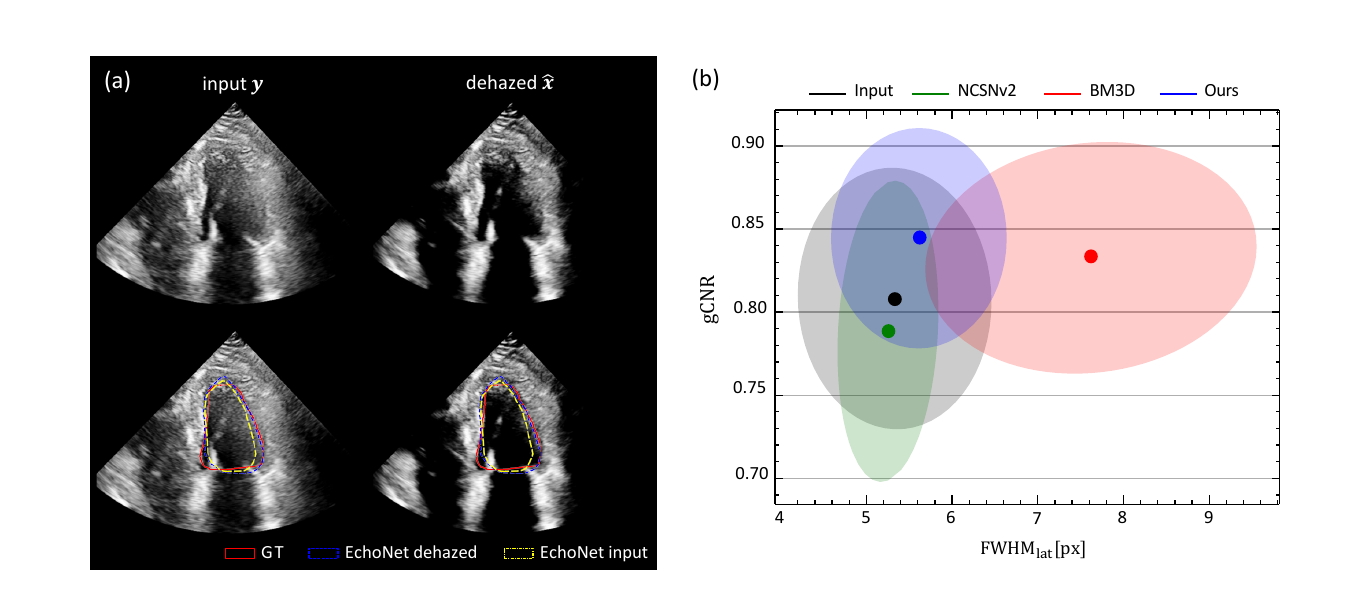}
    \caption{Example: Multipath haze suppression using diffusion models. (a) Example input $\mathbf{y}$ and dehazed output $\hat{\mathbf{x}}$ (top) along with automatic left-ventricular (LV) segmentations by EchoNet \cite{ouyang2019echonet} (bottom). Note how the LV area is underestimated on the hazy, cluttered, input data, and how this is improved after dehazing. (b) Tradeoff between tissue contrast (gCNR) and lateral resolution (FWHM), showing how dehazing by generative diffusion models (ours) greatly improves the gCNR while compromising much less on resolution than denoising methods such as BM3D. It also compares favourably against a discriminative neural network using supervised training on phantom data (NCSNv2). Image adapted from \cite{stevens2024dehazing}.}
    \label{fig:dehazing}
\end{figure*}

\subsection{Active beamsteering using sequential Monte-Carlo}
\subsubsection{Generative model} We start with an agent that is tasked with the sequential selection of optimal focused transmit beams for tracking the position of a moving Doppler target $\mathbf{x}_t= (\theta_t, z_t, \omega_t)$, with $\theta_t$ its angular position, $z_t$ its axial position, and $\omega_t$ its Doppler frequency at timestep $t$ \cite{federici2024active}. { The observation at timestep $t$, $\hat{\mathbf{y}}_t\in \R^{N_\theta}$, is an angular power Doppler profile. It is computed by pixel-based receive beamforming and subsequent Doppler processing of an ensemble of channel data coming from \textit{focused} transmits with a fixed focal depth $z^\textrm{tx}$ and steering angle $\hat{a}_t = \theta^{\textrm{tx}}_t$.} The agent's generative model given a series of actions $a'_{0:T}$ is given by:
\begin{align}
\label{eqn:genmodel_doppler}
p(\mathbf{y}_{0:T}, \mathbf{x}_{0:T} | a'_{0:T}) = p(\mathbf{y}_{0:T} | \mathbf{x}_{0:T}, a'_{0:T})p(\mathbf{x}_{0:T}),
\end{align}
with 
\begin{align}
p(\mathbf{x}_{0:T})=p(\mathbf{x}_0)\prod_{t=1}^T p(\mathbf{x}_t|\mathbf{x}_{t-1}),
\end{align}
assuming $\mathbf{x}$ has sequential Markovian structure, with linear Gaussian state transition dynamics $p(\mathbf{x}_t|\mathbf{x}_{t-1}) = \mathcal{N}(A\mathbf{x}_{t-1},\Sigma_x)$, and a Gaussian memoryless nonlinear observation model $p(y_t|\mathbf{x}_t, a_t) = \mathcal{N}(f(\mathbf{x}_t;a_t),\Sigma_y)$. { Here, $f(\cdot;a_t):\R^{3} \rightarrow \R^{N_\theta}$ is a forward observation function that maps the target state to a simulated power Doppler profile}, using an approximate action-conditional transmit beamprofile.

\vspace{0.3cm}
\subsubsection{Perception}
We here track the posterior using particle filtering \cite{cappe2007overview}, a sequential Monte-Carlo method based on the posterior family in Eqn.~\eqref{eqn:approx_post_sampling}. We use the state transition prior as the proposal distribution (bootstrap particle filter), i.e. we draw proposals as $\mathbf{x}^i_t\sim p(\mathbf{x}^i_t|\mathbf{x}^i_{t-1})$, and update the weights of the resulting $N_p$ particles by their likelihood given the new observations, i.e.:
\begin{align}
\hat{w}^i_t = w^i_{t-1}p(\hat{\mathbf{y}}_t|\mathbf{x}^i_t,a'_t),
\end{align}
and then normalize such that $w^i_t = \frac{\hat{w}^i_t}{\sum_i \hat{w}_t^i}$. As per common practice, we additionally perform resampling of particles to avoid degeneracy \cite{cappe2007overview}. This posterior update then allows sampling updated futures from $p(\mathbf{y}_{t+1}|\hat{\mathbf{y}}_{0:t},a'_{t+1})$, using the particles $\mathbf{y}^i_{t+1}\sim p(\mathbf{y}_{t+1}|\mathbf{x}^i_{t+1},a'_{t+1})$, with $\mathbf{x}^i_{t+1}\sim p(\mathbf{x}^i_{t+1}|\mathbf{x}^i_{t})$. These hypothetical futures are used to drive action selection.

\vspace{0.3cm}
\subsubsection{Action}
At each timestep $t$, the agent greedily selects the next beamsteering action $a_{t+1}^*$ { by maximizing the expected information gain $I(\mathbf{y}_{t+1},\mathbf{x}_{t+1}|a'_{t+1}, \hat{\mathbf{y}}_{0:t})$ across a discrete set of candidate actions $a'_{t+1}\in\mathcal{S}_{a}$, which in light of the aforementioned generative model is equivalent to maximizing the marginal differential entropy of the observations: the conditional entropy is constant for all actions, and solely determined by the receiver noise covariance $\Sigma_y$.} We further assume a marginal multivariate Gaussian distribution and approximate the differential entropy using $N_p$ samples from the marginal $\{\mathbf{y}^1_{t+1}, ..., \mathbf{y}_{t+1}^{N_p}\} \sim p(\mathbf{y}_{t+1}|\hat{\mathbf{y}}_{0:t}, a'_{t+1})$ as:
\begin{align}
a_{t+1}^* =& \underset{a'_{t+1}\in\mathcal{S}_{a}}{\arg\max} \hspace{5pt} \frac{1}{2}\log |\tilde{\Sigma}(a'_{t+1},\hat{\mathbf{y}}_{0:t})|.
\label{eqn:mutual_particlefilter}
\end{align}
This action yields a new observation $\hat{\mathbf{y}}_{t+1}$, which in turn prompts an update of the posterior distribution (perception), closing the loop.

\vspace{0.3cm}
\subsubsection{Results}
Figure~\ref{fig:fetal} shows a real-time implementation of the perception-action loop for a dynamic Doppler target mimicking a beating fetal heart in a lab setup. We use an s5-1 phased array transducer (Philips) mounted on a motion stage, and connected to the Verasonics Vantage 256 system. At each timestep $t$, we perform a Doppler transmit sequence (ensemble length: 2000 frames, PRF: 1.5~kHz, center frequency: 3.125~MHz) with an adaptively-steered focused beam at $\theta_t^{\textrm{tx}}$. After pixel-based receive beamforming, we compute the integral across the depth and Doppler axes to get an observed power-Doppler angular profile $\hat{\mathbf{y}}_t$.  { The information-seeking agent accurately tracks the moving target and retains precise downstream heart-rate estimates by adequately steering the beam and maintaining high Doppler SNR. Performance was also quantified for various Doppler SNR levels \textit{in-silico}, with SNR defined relative to the peak Doppler signal under ideal steering. These results show that heart rate estimation using cognitive adaptive steering remained accurate (within 5 bpm of the ground truth) at about 20dB lower SNR levels compared to non-adaptive steering, a direct result of the dramatic reduction in beamforming gain when the target moved out of the static beam.}

\begin{figure*}
    \centering
    \includegraphics[width=1\linewidth, trim={0.2cm 11cm 10.7cm 0},clip]{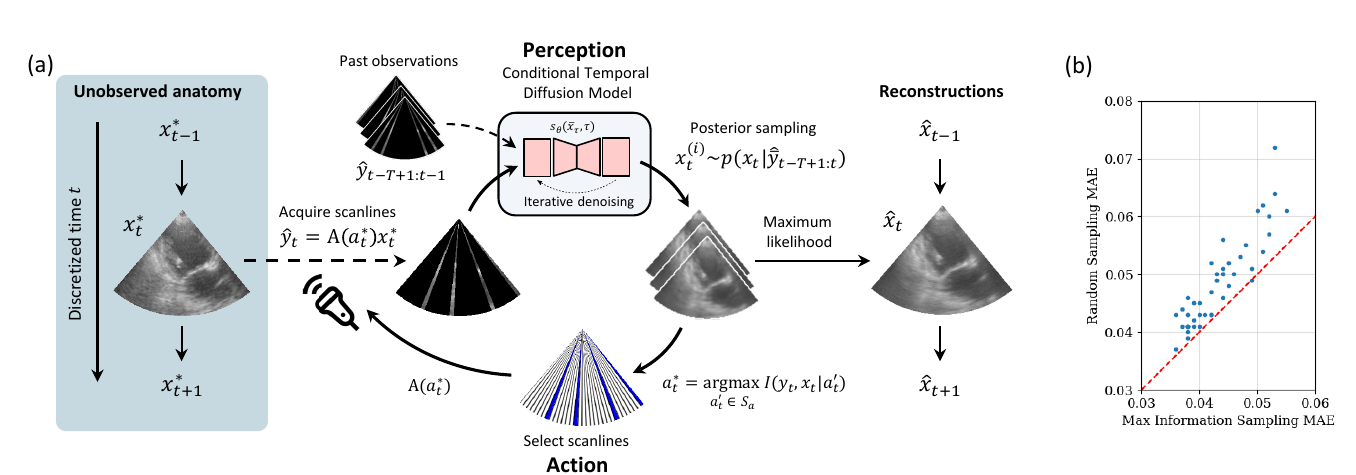}
    \caption{Example: Active subsampling using temporal diffusion models. (a) Overview of the perception-action loop for ultrasound scanline selection. At timestep $t$, the agent acquires $k$ scanlines that maximize expected information gain. It then combines them with the past $k(T-1)$ scanlines acquired at timesteps $t-T+1:t-1$, and perform perceptual inference via diffusion posterior sampling to yield samples $\bx_t^{(i)}$. The most likely sample is selected as the final reconstruction. Next, the posterior samples are used to estimate expected information gain at timestep $t+1$, and the action $a^*_{t+1}$ that maximizes it is used to acquire the next scanlines. (b) Mean absolute reconstruction error (MAE) of a cognitive agent (Max information Sampling) vs a random agent (Random Sampling). Both use the conditional temporal diffusion model for inference. Each blue dot is a hold-out test sequence from the CAMUS dataset. Points above the red dashed line are points for which cognitive imaging outperforms a random scanline selection.}
    \label{fig:diffusion}
\end{figure*}

\subsection{Multipath haze suppression using diffusion models}
\subsubsection{Generative model} The previous example shows how cognitive ultrasound imaging improves target tracking given a physics-based model of the observations and a linear Gaussian state transition function in $\R^3$. To move to more complex and high-dimensional states, such as high-resolution reflectivity maps in $\R^{N\times N}$, we will first need to expand the modelling. { We will therefore train a generative diffusion model from data.}

We consider an agent that is tasked with the suppression of multipath clutter (``haze'') from (linearly) beamformed RF data patches $\mathbf{y}\in\R^{N\times N}$ in cardiac imaging \cite{stevens2024dehazing}. To that end, we define the following generative model, assuming linear scattering:
\begin{align}
\label{eqn:genmodel_haze}
p(\mathbf{y}, \mathbf{x}, \mathbf{h}) = p(\mathbf{y} | \mathbf{x}, \mathbf{h})p(\mathbf{x})p(\mathbf{h}),
\end{align}
where $\mathbf{x}\in\R^{N\times N}$ is the beamformed direct path contribution, and $\mathbf{h}\in\R^{N\times N}$ is the multipath component. The generative model assumes that the direct and multipath RF data are statistically independent, governed by their respective priors $p(\mathbf{x})$ and $p(\mathbf{h})$. We further assume that $p(\mathbf{y} | \mathbf{x}, \mathbf{h})$ is a linear Gaussian model of the form $\mathcal{N}(\mathbf{x}+\mathbf{h},\sigma_n^2 I)$.
We perform denoising score matching on (unpaired) training data samples $\{\hat{\mathbf{x}}^{1}, ..., \hat{\mathbf{x}}^{L}\}$, and $\{\hat{\mathbf{h}}^{1}, ..., \hat{\mathbf{h}}^{L}\}$ to learn two score functions, conditioned on the SDE timestep $\ts$: $\nabla_{\mathbf{x}_\ts}\log p_\tau(\mathbf{x}_\ts) \approx s_\theta(\mathbf{x}_\ts,\ts)$ and $\nabla_{\mathbf{h}_\ts}\log p_\ts(\mathbf{h}_\ts) \approx s_\phi(\mathbf{h}_\ts,\ts)$. For architectural and training details see \cite{stevens2024dehazing}.

\vspace{0.3cm}
\subsubsection{Perception} 
We then perform diffusion posterior sampling $\bx_\ts, \bh_\ts \sim p(\bx_\ts, \bh_\ts|\by)$ through the formulation of a \emph{joint conditional} diffusion process $\left\{\bx_\ts, \bh_\ts|\by\right\}_{\ts\in[0,\mathcal{T}]}$, in turn producing a joint conditional \emph{reverse-time} SDE:
\begin{align}
    \mathrm{d}(\bx_\ts, \bh_\ts) &= 
    \big[
        f(\ts)(\bx_\ts, \bh_\ts) - \ldots \nonumber\\ &g(\ts)^2 
            \nabla_{\bx_\ts, \bh_\ts} \log{p(\bx_\ts, \bh_\ts|\by})
    \big] \mathrm{d}\tau + g(\ts) \mathrm{d}\Bar{\mathbf{w}}_\ts,
\label{eq:cond_reverse_diff}
\end{align}
where the posterior score at SDE timestep $\ts$, $\nabla_{\bx_\ts, \bh_\ts} \log{p(\bx_\ts, \bh_\ts|\by})$, is approximated as:
\begin{equation}
\label{eq:approx_cond}
    \left\{
        \begin{array}{lr}
                \nabla_{\bx_\ts}\log{p(\bx_\ts, \bh_\ts|\by)} \simeq s_\theta(\bx_\ts, \ts) \\ \hspace{4.5cm} + \nabla_{\bx_\ts} \log{p(\by|\bx_\ts, \bh_\ts)}
                \\[4pt]
                \nabla_{\bh_\ts}\log{p(\bx_\ts, \bh_\ts|\by)} \simeq s_\phi(\bh_\ts, \ts) \\ \hspace{4.5cm} + \nabla_{\bh_\ts} \log{p(\by|\bx_\ts, \bh_\ts)}
        \end{array},
    \right.
\vspace{-0em}
\end{equation}
in which we use the generative model in Eqn.~\eqref{eqn:genmodel_haze}, Bayes rule, and that the gradient of the marginal $p(\mathbf{y})$ with respect to $\mathbf{x}$ and $\mathbf{h}$ is zero. As noted in section~\ref{sec:DPS}, the conditional distribution of $\by$ given noise-perturbed states $\bx_\ts$ and $\bh_\ts$, $p(\by|\bx_\ts, \bh_\ts)$, is generally intractable, unlike the known data likelihood factor in the generative model $p(\by|\bx_0, \bh_0)$. Following Song \textit{et al}. \cite{songScoreBasedGenerativeModeling2021}, we therefore approximate it as $p(\by|\bx_\ts, \bh_\ts) \approx p(\tilde{\by}_\ts|\bx_\ts, \bh_\ts) = \mathcal{N}(\bx_\ts + \bh_\ts, \rho^2 \bI)$, with $\tilde{\mathbf{y}}_\ts\sim q(\mathbf{y}_\ts|\mathbf{y}_0)$ being a noise-perturbed observation. 

\vspace{0.3cm}
\subsubsection{Results} 
We evaluate performance on 1020 cardiac ultrasound RF frames from two difficult-to-image volunteers across 4 cardiac views (three chamber, four chamber, parasternal long axis and parasternal short axis), acquired using an X5-1C matrix transducer connected using a Philips EPIQ scanner in harmonic imaging mode. Figure~\ref{fig:dehazing}a qualitatively shows how dehazing significantly boosts tissue contrast, as well as improving the downstream left-ventricular segmentation. In Fig.~\ref{fig:dehazing}b, we quantitatively evaluate performance, showing that our deep generative dehazing model strikes a good balance between tissue contrast and resolution, comparing favourably to other methods.

\subsection{Active subsampling using temporal diffusion models}
\subsubsection{Generative Model}
{ Our final example concerns an agent that performs adaptive compressive sensing \cite{van2021active, braun2015info} through active subsampling of ultrasound scanlines. We will use the approach by Nolan \textit{et al.} \cite{nolan2024active} to design subsampling masks that maximize information gain given a diffusion-based generative model.} Subsampling transmit events (be it scanlines, diverging waves, or otherwise) is typically used to minimize costs associated with data acquisition, such as acquisition time. Our agent is tasked with the reconstruction of a sequence of $T$ ultrasound frames $\bx_{1:T}\in \R^{N_z \times N_y \times T}$ from goal-directed partial observations $\by_{1:T}\in \R^{N_z \times k \times T}$, with a budget of $k<N_y$ focused scanlines per frame. We formulate the following observation model:
\begin{equation}
\by_t=\mathrm{A}(a_t)\bx_t,
\end{equation}
with $\mathrm{A}(a_t)$ being the time-varying (structured) subsampling matrix that given action $a_t$ selects all samples belonging to $K$ scanlines from the vectorized frame $\bx_t$. The agent's generative model is then given by:
\begin{align}
\label{eqn:genmodel_diffusion}
p(\mathbf{y}_{1:T}, \mathbf{x}_{1:T} | a'_{1:T}) =& p(\mathbf{y}_{1:T} | \mathbf{x}_{1:T}, a'_{1:T})p(\mathbf{x}_{1:T}), \\
 =& p(\mathbf{x}_{1:T})\prod_{t=1}^T p(\mathbf{y}_{t} | \mathbf{x}_{t}, a'_t),
 \label{eqn:diffusion_factorized}
\end{align}
assuming a Gaussian memoryless linear observation model $p(\by_t|\mathbf{x}_t, a'_t) = \mathcal{N}(\mathrm{A}(a'_t),\Sigma_y)$ that can be factorized in time.

We will again use a diffusion model to establish the spatiotemporal generative prior $p(\bx_{1:T})$. To that end, we sample sequences of $T=4$ frames $\hat{\bbx}_t^l=[\hat{\bx}^l_{t-T+1},...,\hat{\bx}^l_{t}]$ from the CAMUS dataset \cite{leclerc2019deep}, and convert them into polar coordinates. We use denoising score-matching to train the (now spatiotemporal) score function $s_\theta(\bbx_\ts, \ts) \approx \nabla_{\bbx_\ts}\log p_\ts(\bbx_\ts)$ using a U-Net architecture based on \cite{diffusionkeras}. We use 400 videos for training and 50 for testing. One video contained about 19 frames on average.

\vspace{0.3cm}
\subsubsection{Perception}
At each timestep $t$, we consider the $T$ most recent measurements $\hat{\by}_{t-T+1:t}$, and use diffusion posterior sampling to generate a total of 16 posterior samples $\{\bbx^{(1)}_{t},...,\bbx^{(16)}_{t}\}\sim p(\bbx_{t} | \hat{\by}_{t-T+1:t})$, tracking a distribution over plausible { anatomical state} sequences. To that end, we again exploit Tweedy's formula and perform likelihood-guidance of the reverse diffusion process making use of the factorization in Eqn.~\eqref{eqn:diffusion_factorized}. The posterior sample with the highest likelihood is used to produce reconstruction $\hat{\bx}_t$ which is compared with the ground truth $\bx_t^*$.

\vspace{0.3cm}
\subsubsection{Action}
Using the { generative forward model}, we project the posterior samples $\bx^{(i)}_{t}$ into hypothetical future observations $\by_{t+1}|a'_{t+1}$, and greedily choose the $k$ lines that maximize expected information gain at timestep $t+1$, i.e. $a^*_{t+1} = \underset{a'_{t+1}\in \mathcal{S}_a}{\arg\max} 
\hspace{4pt} I(\mathbf{y}_{t+1},\mathbf{x}_{t+1}|a'_{t+1}, \hat{\mathbf{y}}_{t-T+1:t})$, using a pixel-wise Gaussian marginal entropy model. { As in the first example, the conditional entropy is constant across the actions, and solely determined by $\Sigma_y.$ In effect, the agent will thus select those lines that have the highest generalized variance across the posterior samples, in order to resolve that uncertainty. } Figure~\ref{fig:diffusion}a gives an overview of the approach.

\vspace{0.3cm}
\subsubsection{Results} The model successfully tracks the anatomical dynamics over time and { generates accurate reconstructions } preserving important anatomical details using only 12.5\% of the scan lines per frame. We compare the cognitive sampling strategy to random sampling and find that maximum information sampling almost always outperforms random sampling (see Figure~\ref{fig:diffusion}b). We also find that maximum information sampling with 4 lines outperforms random sampling with 6 lines, indicating that cognitive sampling, in this case, is worth 50\% more measurements.

\section{Discussion}
\label{sec:discussion}
\noindent This paper gives a new interpretation of the ultrasound transmit-receive cycle as a perception-action loop, in which pulse-echo experiments are \textit{adaptively designed} to maximize information gain given a generative model. It treats ultrasound imaging as an active inference problem \cite{friston2016active}, with clear ties to other fields that have a long tradition, such as Bayesian experiment design \cite{chaloner1995bayesian}, active perception \cite{bajcsy1988active}, active hypothesis testing \cite{naghshvar2013active}, and adaptive compressed sensing \cite{braun2015info}. We postulate that recent advances in deep generative modelling unlock the potential of these approaches for high-dimensional imaging \cite{nolan2024active, van2023active}, and specifically ultrasound.

{ Active inference also shares similarities with other fields, such as reinforcement learning (RL) and stochastic optimal control \cite{rawlik2013stochastic}. In stochastic optimal control, policies are designed to minimize an expected KL divergence to a goal prior. Like in active inference, actions are selected based on real-time simulations (inference) of future states based on past and present observations. In both active inference and stochastic optimal control, an optimal policy can be specified as some explicit functional of a probability density function about the expected future, e.g. an explicit information measure (such as the mutual information), a free energy, or a divergence such as the KL divergence. In deep RL, optimized policies are derived from data based on (delayed) rewards computed using simulations of actions and resulting observations (episodes). In some variants of RL, it is the action-value function that is learned from data (Q learning), and the policy is to execute the action that maximizes this action-value function \cite{hester2018deep}.}

{ The examples given in this paper illustrate the concept of cognitive ultrasound, but also have limitations. In example VII-C, we considered focused adaptive beamsteering actions that lead to individual beamformed lines, while improved resolution could be achieved using software-based retrospective transmit beamforming. For instance, transmit sequences can be designed to optimize recovery of the full multistatic dataset that is in turn used for retrospective transmit beamforming \cite{spainhour2024optimization}.} Moreover, the cognitive ultrasound paradigm extends towards optimization of other transmit parameters, including the waveform. Its efficacy hinges upon an accurate and sufficiently fast model that predicts the consequences of changing such parameters on the received channel data, based on all the channel data received in the past. Neural approximations of physics-based acoustics simulators may allow flexible trade-offs between inference speed and accuracy.  

Using cognitive imaging, a system might in the future perform very resource-efficient 3D scanning, e.g. using combinations of focused and unfocused transmissions in 3D that jointly maximise the information gain. Moreover, a cognitive ultrasound system might optimize the full transmitted wavefield to minimize the impact of the acoustic window and other patient-specific challenges in difficult-to-image patients. This again requires modeling and inclusion of multipath effects. Multipath could either be modeled as stochastic structured noise (as we did for the de-hazing example), or as a deterministic element of the forward model. In the stochastic setting, a cognitive system could e.g. strive to shape the statistics of the multipath signal such that it is easily separated from the direct path.

While we see a lot of opportunities and possible applications, there are also many open questions and challenges. First, using deep generative models introduces a bias that is highly desired (the structure of the natural world), but it potentially also introduces undesired biases e.g. arising from inadequate or biased sampling of the true data distribution when training. Care must be taken not to make priors too informative and restrictive. Second, accurate posterior estimation in complex models comes with high computational complexity, and balancing this accuracy with efficiency for real-time implementations is not trivial, potentially requiring hierarchical inference with adaptive complexity. 
Third, inferring sequences of future actions across longer time horizons (instead of our current greedy approach) is computationally daunting, with naive implementations growing exponentially in complexity with time. This will likely require clever sequential reduction of the action space. Future work should be geared towards addressing these challenges.

\section{Conclusion}
\label{sec:conclusion}
\noindent Ultrasound systems engage in repeated reciprocal interactions with the anatomical environment that they probe, and this cycle of probing and receiving data can be interpreted as a perception-action loop. Active inference offers a probabilistic framework for developing agents that implement such perception-action loops, which when augmented with deep generative models that govern anatomical beliefs, unlocks closed-loop high-dimensional imaging. We here showed how these principles can be used for cognitive ultrasound systems that maximize information gain by changing their probing of the environment. The promise that this holds for ultrasound imaging is significant; it may spur a paradigm shift in the design of systems, akin to cognitive radar systems, that autonomously strive to maximize diagnostic value in-situ.

\section*{Acknowledgment}
\noindent The author would like to thank Tristan Stevens, Oisín Nolan, Wessel van Nierop, and Beatrice Federici, for their contributions in particular to section~\ref{sec:examples}. 

\begin{IEEEbiography}[{\includegraphics[width=1in,height=1in,clip,keepaspectratio]{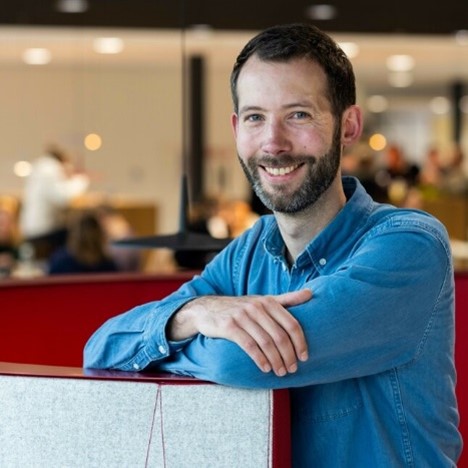}}]{Ruud J.G. van Sloun} is an Associate Professor at the Department of Electrical Engineering at the Eindhoven University of Technology. He received the M.Sc. and Ph.D. degree (both cum laude) in Electrical Engineering from the Eindhoven University of Technology, Eindhoven, The Netherlands, in 2014, and 2018, respectively. From 2019-2020 he was a Visiting Professor with the Department of Mathematics and Computer Science at the Weizmann Institute of Science, Rehovot, Israel, and from 2020-2023 he was a kickstart AI fellow at Philips Research. He received an ERC starting grant, an NWO VIDI grant, an NWO Rubicon grant, and a Google Faculty Research Award. His current research interests include closed-loop image formation, deep generative learning for signal processing and imaging, active signal acquisition, model-based deep learning, compressed sensing, ultrasound imaging, and probabilistic signal and image reconstruction.
\end{IEEEbiography}

\bibliography{bib} 
\bibliographystyle{ieeetr}

\end{document}